\documentclass[ twocolumn]{aastex631}

\usepackage{xcolor}
\usepackage[utf8]{inputenc}

\usepackage{soul}
\shorttitle{AMEGO-X single-site events}
\shortauthors{AMEGO-X}
\graphicspath{{./}{figures/}}
\begin{document}

\title{Improving the Low-Energy Transient Sensitivity of AMEGO-X using Single-Site Events}

\author[0000-0002-2471-8696]{I. Martinez-Castellanos}
\affiliation{NASA Goddard Space Flight Center, Greenbelt, MD 20771}
\affiliation{Department of Astronomy, University of Maryland, College Park, Maryland 20742, USA}
\affiliation{Center for Research and Exploration in Space Science and Technology, NASA/GSFC, Greenbelt, Maryland 20771, USA}

\author[0000-0002-0794-8780]{Henrike Fleischhack}
\affiliation{Catholic University of America, 620 Michigan Ave NE, Washington, DC 20064, USA}
\affiliation{NASA Goddard Space Flight Center, Greenbelt, MD 20771}
\affiliation{Center for Research and Exploration in Space Science and Technology, NASA/GSFC, Greenbelt, Maryland 20771, USA}

\author[0000-0002-6774-3111]{C. Karwin}
\affiliation{Department of Physics and Astronomy, Clemson University, Clemson, SC 29634, USA}

\author[0000-0002-6548-5622]{M. Negro}
\affiliation{Center for Space Sciences and Technology, University of Maryland, Baltimore County, Baltimore, MD 21250, USA}
\affiliation{NASA Goddard Space Flight Center, Greenbelt, MD 20771}
\affiliation{Center for Research and Exploration in Space Science and Technology, NASA/GSFC, Greenbelt, Maryland 20771, USA}

\author[0000-0002-9852-2469]{D. Tak}
\affiliation{Deutsches Elektronen-Synchrotron (DESY), Platanenallee 6, Zeuthen, 15738, Germany}

\author[0000-0002-7851-9756]{Amy Lien}
\affiliation{University of Tampa, Department of Chemistry, Biochemistry, and Physics, 401 W. Kennedy Blvd, Tampa, FL 33606, USA}

\author[0000-0001-6677-914X]{C. A. Kierans}
\affiliation{NASA Goddard Space Flight Center, Greenbelt, MD 20771}

\author[0000-0002-9249-0515]{Zorawar Wadiasingh}
\affiliation{Department of Astronomy, University of Maryland, College Park, Maryland 20742, USA}
\affiliation{NASA Goddard Space Flight Center, Greenbelt, MD 20771}
\affiliation{Center for Research and Exploration in Space Science and Technology, NASA/GSFC, Greenbelt, Maryland 20771, USA}

\author[0000-0002-0921-8837]{Yasushi Fukazawa}
\affiliation{Department of Physics, Hiroshima University,
1-3-1 Kagamiyama, Higashi-Hiroshima, Hiroshima, Japan, 739-8526}

\author[0000-0002-6584-1703]{Marco Ajello}
\affiliation{Department of Physics and Astronomy, Clemson University, Clemson, SC 29634, USA}

\author[0000-0003-4433-1365]{Matthew~G.~Baring}
\affiliation{Department of Physics and Astronomy - MS 108, Rice 
University, 6100 Main Street, Houston, Texas 77251-1892, USA}

\author[0000-0002-2942-3379]{E. Burns}
\affiliation{Department of Physics \& Astronomy, Louisiana State University, Baton Rouge, LA 70803, USA}

\author[0000-0002-9280-836X]{R. Caputo}
\affiliation{NASA Goddard Space Flight Center, Greenbelt, MD 20771}

\author[0000-0002-8028-0991]{Dieter H. Hartmann}
\affiliation{Department of Physics and Astronomy, Clemson University, Clemson, SC 29634, USA}

\author[0000-0001-9608-4023]{Jeremy S. Perkins}
\affiliation{NASA Goddard Space Flight Center, Greenbelt, MD 20771}

\author[0000-0002-4744-9898]{Judith L. Racusin}
\affiliation{NASA Goddard Space Flight Center, Greenbelt, MD 20771}

\author[0000-0002-3833-1054]{Yong Sheng}
\affiliation{Department of Physics and Astronomy, Clemson University, Clemson, SC 29634, USA}

%% Note that the \and command from previous versions of AASTeX is now
%% depreciated in this version as it is no longer necessary. AASTeX 
%% automatically takes care of all commas and "and"s between authors names.

%% AASTeX 6.31 has the new \collaboration and \nocollaboration commands to
%% provide the collaboration status of a group of authors. These commands 
%% can be used either before or after the list of corresponding authors. The
%% argument for \collaboration is the collaboration identifier. Authors are
%% encouraged to surround collaboration identifiers with ()s. The 
%% \nocollaboration command takes no argument and exists to indicate that
%% the nearby authors are not part of surrounding collaborations.

%% Mark off the abstract in the ``abstract'' environment. 
\begin{abstract}

AMEGO-X, the All-sky Medium Energy Gamma-Ray Observatory eXplorer, is a proposed instrument designed to bridge the so-called ``MeV gap'' by surveying the sky with unprecedented sensitivity from $\sim$100 keV to about one GeV. This energy band is of key importance for multi-messenger and multi-wavelength studies but it is nevertheless currently under-explored. AMEGO-X addresses this situation by proposing a design capable of detecting and imaging gamma rays via both Compton interactions and pair production processes. However, some of the objects that AMEGO-X will study, such as gamma-ray bursts and magnetars, extend to energies below $\sim$100 keV where the dominant interaction becomes photoelectric absorption. These events deposit their energy in a single pixel of the detector. In this work we show how the $\sim$3500 cm$^{2}$ effective area of the AMEGO-X tracker to events between $\sim$25 keV to $\sim$100 keV will be utilized to significantly improve its sensitivity and expand the energy range for transient phenomena. Although imaging is not possible for single-site events, we show how we will localize a transient source in the sky using their aggregate signal to within a few degrees. This technique will more than double the number of cosmological gamma-ray bursts seen by AMEGO-X, allow us to detect and resolve the pulsating tails of extragalactic magnetar giant flares, and increase the number of detected less-energetic magnetar bursts ---some possibly associated with fast radio bursts. Overall, single-site events will increase the sensitive energy range, expand the science program, and promptly alert the community of fainter transient events.

\end{abstract}

%% Keywords should appear after the \end{abstract} command. 
%% The AAS Journals now uses Unified Astronomy Thesaurus concepts:
%% https://astrothesaurus.org
%% You will be asked to selected these concepts during the submission process
%% but this old "keyword" functionality is maintained in case authors want
%% to include these concepts in their preprints.
\keywords{Gamma-ray astronomy (628) --- Gamma-ray detectors (630) --- Time domain astronomy (2109) --- Gamma-ray bursts (629) --- Magnetars (922)}

%% From the front matter, we move on to the body of the paper.
%% Sections are demarcated by \section and \subsection, respectively.
%% Observe the use of the LaTeX \label
%% command after the \subsection to give a symbolic KEY to the
%% subsection for cross-referencing in a \ref command.
%% You can use LaTeX's \ref and \label commands to keep track of
%% cross-references to sections, equations, tables, and figures.
%% That way, if you change the order of any elements, LaTeX will
%% automatically renumber them.
%%
%% We recommend that authors also use the natbib \citep
%% and \citet commands to identify citations.  The citations are
%% tied to the reference list via symbolic KEYs. The KEY corresponds
%% to the KEY in the \bibitem in the reference list below. 

\section{Introduction} \label{sec:intro}

Gamma-ray observations have played a central role in multi-messenger astronomy. The detection of the gamma-ray burst GRB 170817A by \textit{Fermi}-GBM shortly after the gravitational wave signal GW170817 \citep{grbgw170817}, and the gamma-ray flare from the blazar TXS 0506+056 seen by \textit{Fermi}-LAT in coincidence the high-energy neutrino IceCube-170922A \citep{TXSmulti2018} resulted in unique insights which would not have been possible otherwise. In the near future, gamma-ray transients will also be critical for attaining counterparts to gravitational wave signals and understanding the details of neutron star mergers \citep{2019BAAS...51c.260B}. Gamma-ray observations also provide critical information needed to understand the physics of cosmic ray acceleration and the sources that produce these energetic charges \citep{10.1093/mnras/stx3280, 10.1093/mnras/stx498,2019BAAS...51c.431O,2019BAAS...51c.151O,2019BAAS...51c.396V,2019BAAS...51c.485V}.

Despite this success, there is a portion of the gamma-ray spectrum that remains relatively unexplored. The so-called ``MeV gap'', from approximately 0.1 MeV to 100 MeV, has not been studied with the same sensitivity as the neighboring energies, lagging behind by more than order of magnitude in flux. There is also great discovery potential in this energy range, in part due to the fact that some classes of AGN \citep[e.g.,][]{1998MNRAS.299..433F,2019arXiv190306106P,2019BAAS...51c.485V,2019BAAS...51c.348R,2019BAAS...51c.291M}, magnetars in quiescence  \citep[e.g.,][]{2008A&ARv..15..225M,Turolla2015,2019BAAS...51c.292W}, rotation-powered pulsars \citep[e.g.,][]{2015MNRAS.449.3827K,2019BAAS...51c.379H} and topical systems involving pulsar winds and leptonic pevatrons \citep{1996ApJ...457..253D,2013A&ARv..21...64D,2013arXiv1305.2552K,2015SSRv..191..391K,2017hsn..book.2159S,2017ApJ...850..100A,2019BAAS...51c.513G,2019BAAS...51c.183D,2020ApJ...904...91V,2020ApJ...897...52A,2021MNRAS.502..915C,2021arXiv210801705W,2021arXiv211102575A} have a peak energy output in the 0.1-100 MeV band. Moreover, MeV observations can be constraining to some models and sectors of particle dark matter \citep[e.g.,][]{2019BAAS...51c..78C,2021JCAP...06..036F}.

%\zw{Should there be a few sentences on how AMEGO-X is different from AMEGO?} IMC: Added a sentence and reference.
AMEGO-X\footnote{\url{https://asd.gsfc.nasa.gov/amego-x}} is a proposed mission capable of surveying the sky with unprecedented sensitivity at these energies \citep{2021arXiv210802860F}, more than an order of magnitude better compared to its predecessors \citep{comptel1985sensi}. As explained in Section \ref{sec:mission}, AMEGO-X is equipped with a silicon pixel tracker and a cesium iodide scintillator calorimeter, tuned to detect gamma-ray photons both through Compton scattering interactions ($\sim$100 keV to $\sim$10 MeV) and electron-positron pair production ($\sim$10 MeV to $\sim$1 GeV). Two or more hits on either the tracker or the calorimeter allow for distinguishing between these two types of events, and for the reconstruction of the direction and energy of the incident photon. AMEGO-X builds upon the AMEGO concept \citep{McEnery2019All}, with a reduced energy threshold, improved sensitivity at $\sim$GeV energies, and a smaller and lighter detector.

The science enabled by AMEGO-X is vast and varied. While the measurements in the MeV band will be crucial, some of the sources of interest have a spectral distribution that extends below the energy threshold for Compton events. As discussed in Section \ref{sec:science}, the energy output of some of these sources peaks below 200 keV. This is the case of some gamma-ray burst (GRBs), magnetar giant flare (MGFs) tails, and recurrent magnetar short bursts ---such as those associated with a fast radio burst in SGR 1935+2154 \citep{gbm_catalog_spectra_2014,2015ApJS..218...11C,2020ApJ...898L..29M,Bochenek_2020,2020Natur.587...54C,Burns2021,2021NatureSvinkin,hurley2005exceptionally, 2021NatAs...5..378L,2021NatAs...5..408Y,1999Natur.397...41H,1999AstL...25..635M,1999ApJ...515L...9F}. Extending the energy range of AMEGO-X to lower energies will significantly enhance its capacity to study these different source classes.

In this work we demonstrate that it is possible to extend the energy range of AMEGO-X down to 25 keV using \textit{single-site events} (SSE). These are events that produce a single hit, in most cases due to photoelectric absorption. The cross section for this type of interaction increases rapidly below $\sim$200 keV.

We show in Section \ref{sec:performance} how the large effective area between 25 keV and 100 keV allows us to achieve a competitive performance for transient events of up to 100 s duration. Moreover, despite the fact that, by definition, SSE leave no tracks and therefore event-by-event direction reconstruction is not possible, we show how the spatial distribution of the SSE in the detector can be used to constrain the location of a burst like GRB 170817A to within a $<$2$^{\circ}$ radius.

Leveraging the information from SSE will enhance the AMEGO-X capabilities overall. It will enlarge the observable volume and increase the rate of detected GRBs and other transients, supporting the multi-wavelength and multi-messenger efforts of the astronomy and astroparticle communities. We will be able to measure energy spectra down to tens of keV, improving our ability to reject or support theoretical models.  This spectral extension will assist AMEGO-X in its goal of significantly improving our knowledge of transient sources in the MeV band.

\section{The AMEGO-X mission}
\label{sec:mission}

\begin{figure*}[tb]
    \centering
    \gridline{\fig{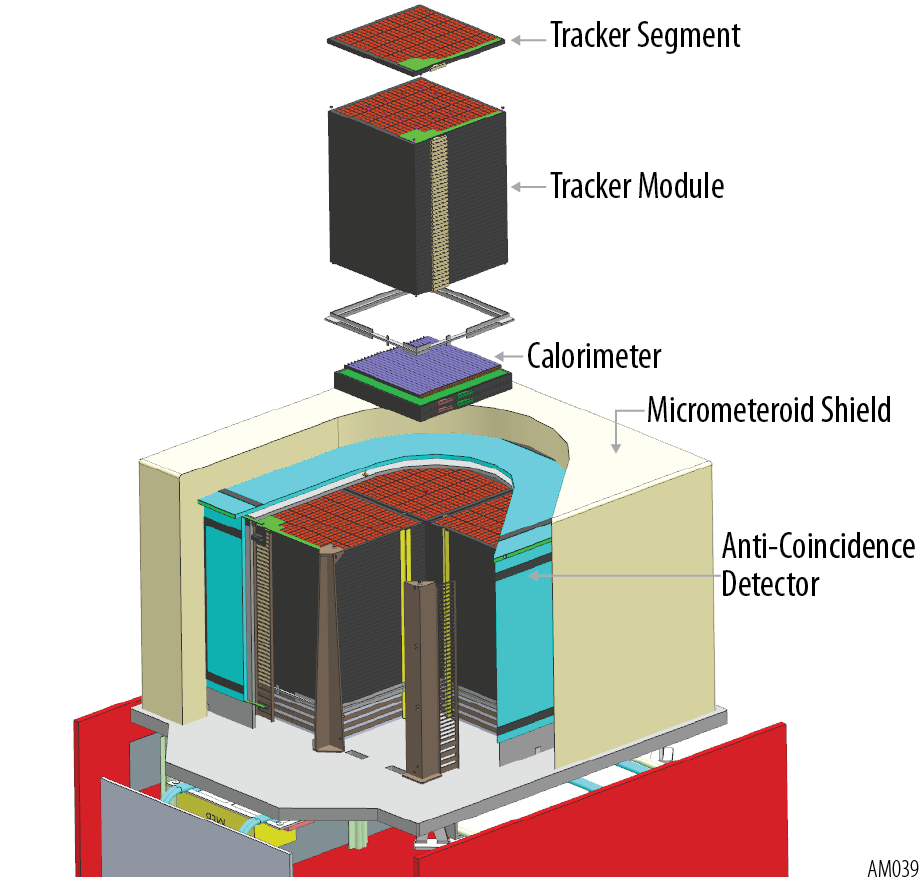}{0.45\textwidth}{(a)}
	              \fig{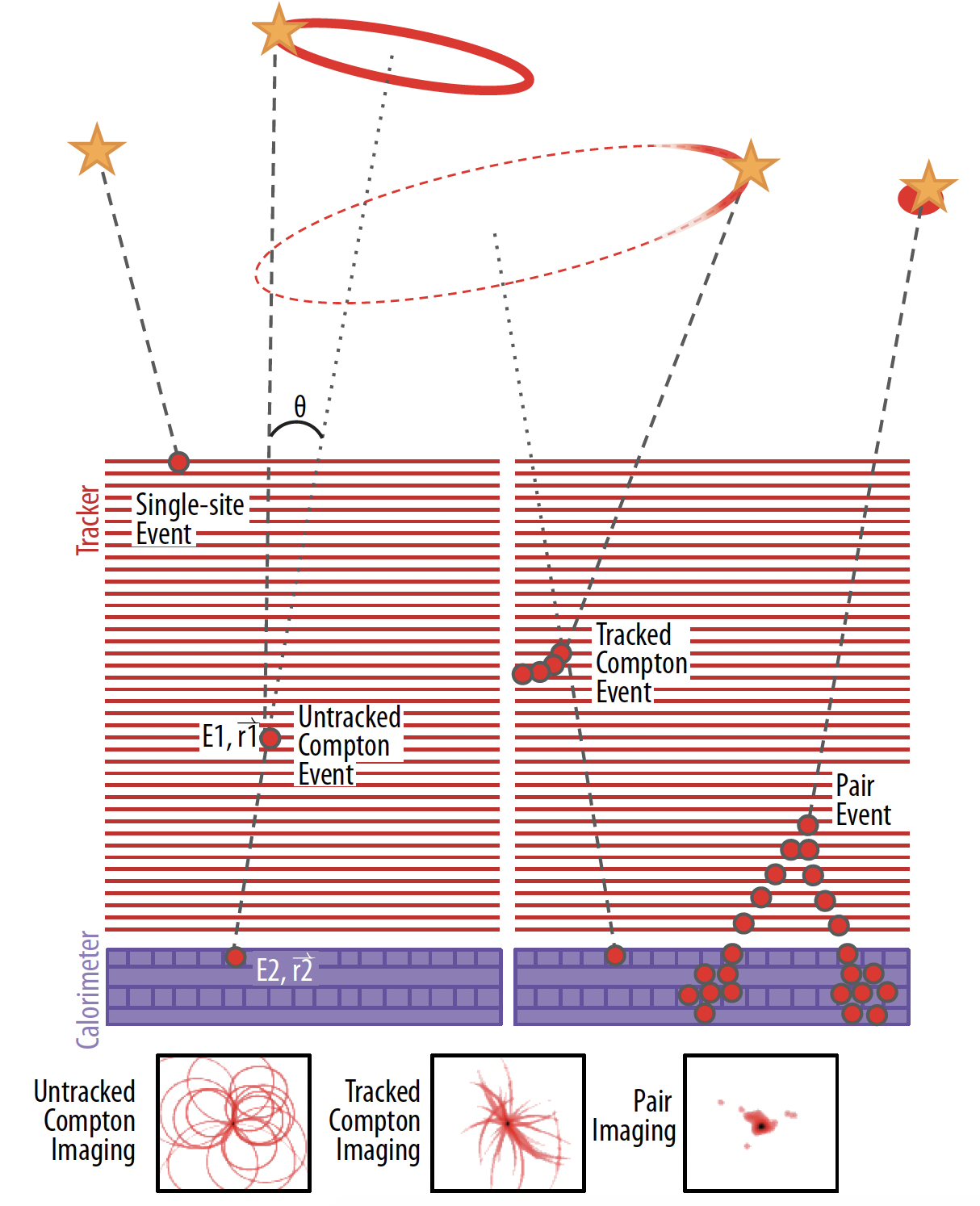}{0.35\textwidth}{(b)}}
    \caption{(a) The AMEGO-X Gamma-Ray Telescope has four identical Detection Towers, each composed of a 40 layer silicon pixel Tracker and 4 layer CsI Calorimeter. The Detection Towers are surrounded by an ACD and Micrometeoroid Shield. (b) AMEGO-X is designed to detect Compton scattering and pair production events, where the position and energy deposited in multiple interactions within the large detector volume are used to reconstruct the direction of the incoming gamma ray. Single site events, which result from mainly photoelectric absorption within a single pixel, cannot be used for single-photon imaging, but allow for increased low-energy sensitivity to transient events.}
    \label{fig:amego-xschematic}
\end{figure*}

The AMEGO-X Gamma-Ray Telescope  (GRT) consists of a silicon pixel Tracker, a hodoscopic Cesium Iodide (CsI) Calorimeter, surrounded by a plastic anti-coincidence detector (ACD), as shown in Figure~\ref{fig:amego-xschematic}~(a). 
The GRT consists of four identical detection towers each with a Tracker and Calorimeter Module.
The Tracker Module has 40 layers, each measuring $40\times40$~cm$^2$ and separated by 1.5~cm, composed of monolithic complementary metal-oxide-semiconductor (CMOS) active pixel sensors (APS). The Calorimeter consists of 4 layers, each with 25 $1.5\times1.5\times38$~mm$^3$ CsI bars, modelled after the \textit{Fermi}-LAT calorimeter. 

To operate as a telescope sensitive to Compton scattering and pair production interactions across four decades of energy, AMEGO-X requires a large instrumented area, with high stopping power, and good position and energy resolution in the sensitive detector volumes. 
The GRT design is similar to \textit{Fermi}-LAT, although it is optimized for the low MeV range with a 3D position-sensitive Tracker and does not have conversion foils~\citep{2009ApJ...697.1071A}.
Specifically, the main functionality of the Tracker is to measure the energy and position of gamma-ray Compton scatter and pair-conversion charged-particle interactions with high precision. As seen in Figure~\ref{fig:amego-xschematic}  these type of interactions are denominated \textit{Compton events} ---either tracked (TC) or untracked (UC), depending on whether or not the direction of the scattered electron can be determined--- and \textit{pair events} (P) respectively.
The AMEGO-X instrument is further described in \cite{2021arXiv210802860F}.

AMEGO-X leverages more than 10 years of development in CMOS monolithic APS technology from the ground-based particle physics experiments~\citep{peric2021}. 
The Tracker detectors are based on the AstroPix project \citep{BREWER2021165795} which has optimized the ATLASPix detector design \citep{peric2012} for gamma-ray applications. 
Each AMEGO-X Tracker layer contains an array of 380 AstroPix $2\times2$~cm APS chips, each with $19\times17$ pixels, where the pixel size is $1\times1$~mm$^2$. 
The APS are 0.5 mm thick and operate at full depletion. 
Each interaction within the Tracker is recorded as the energy deposited (25-700 keV) and the pixel location (0.5~mm$^3$ resolution). The low detector threshold of 25 keV allowed by the Tracker APS and the ability to obtain three-dimensional spatial information are two key features of the AstroPix detector that enabled this work.

The main trigger criteria requires at least two detected hits on either the tracker or the calorimeter, resulting from Compton or pair-production interactions. At energies below the Compton regime ($\lesssim$100 keV), photons predominantly undergo photoelectric absorption and their energy is deposited in a single pixel. 
With the high gamma-ray backgrounds at these energies and no imaging capabilities, single-site events (SSE) are dominated by background for most observations and are not continously recorded due to prohibitively high data rates (see Section~\ref{sec:background}). SSE are instead saved into a temporary buffer, and short duration ($\sim$100 seconds) readouts are initiated when the on-board Transient Alert (TA) logic identifies a significant rate increase above the background. 

The AMEGO-X on-board TA logic is based on \textit{Fermi}-GBM algorithms~\citep{paciesas2012}. This capability allows to measure emission down to 25 keV and expands the portfolio of astrophysical phenomena beyond the previously conceived sensitivity to include lower-energy transients, especially GRBs, giant magnetar flares, and magnetar short bursts, as discussed in Section~\ref{sec:science}.
AMEGO-X will send rapid alerts via the Tracking and Data Relay Satellite System (TDRSS) Demand Access within 30 seconds to the Gamma-ray Coordinates Network (GCN) enabling multi-wavelength follow-up.

\section{Single-site events performance}
\label{sec:performance}

The response of the AMEGO-X instrument has been simulated using the Medium-Energy Gamma-ray Astronomy library (MEGAlib) software package\footnote{Available at \url{https://megalibtoolkit.com/home.html}}, a standard tool in MeV astronomy~\citep{2006NewAR..50..629Z}. MEGAlib is based on GEANT4 \citep{geant4_2003} and is able to simulate the emission from a source, the passage of particles through the spacecraft and their energy deposits in the detector. It also performs event reconstruction, imaging analysis and allows to estimate the background rates from different expected components. 

The knowledge of the effective area and the expected background is critical to understand the advantages of utilizing SSE over only considering Compton and pair events. We estimate the sensitivity ratio between these two cases for various spectral hypotheses which leads to the realization that some types of studies are only possible when SSE are considered, as described in Section \ref{sec:science}. 

In this section we also describe the expected localization performance for burst-like events using the aggregate signal from SSE. This will allow AMEGO-X to provide sky localization information ---critical for multi-messenger and multi-wavelength detection--- even for soft-spectrum sources with a low signal in the Compton channel.

\subsection{Effective area}

The ability to perform imaging analysis reduces the background and increases the sensitivity of an instrument. Although this is not an option for SSE, the increase in background with respect to Compton and pair events is well compensated by the large effective area to this type of events. Each tracker layer has a physical area of approximately 6400 cm$^{2}$, and together they efficiently absorb these low-energy photons. 

\begin{figure}
\centering
\includegraphics[width=\columnwidth]{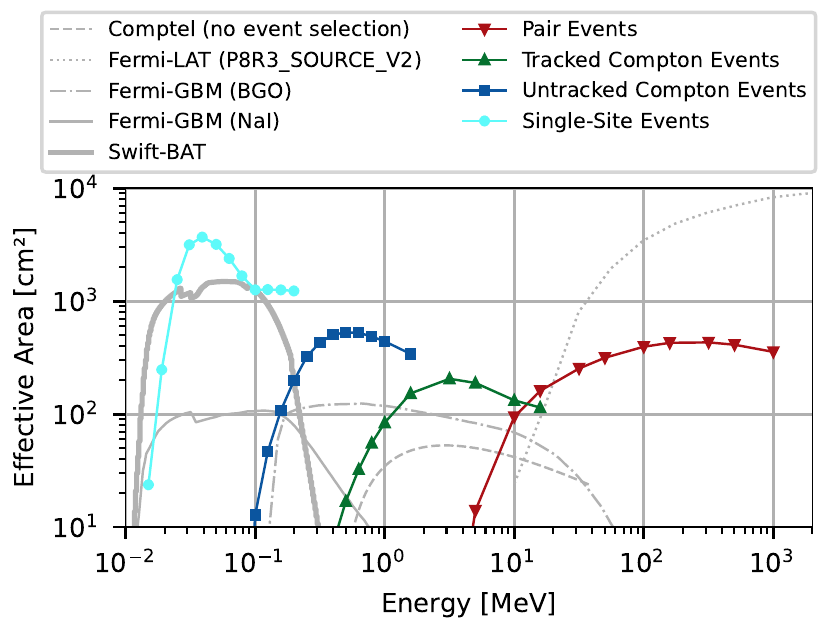}
\caption{Simulated effective area for different event types recorded by AMEGO-X as a function of the photon energy, for incident photons parallel to the boresight (on-axis). The grey lines show the effective areas of other instruments: \emph{Fermi}-GBM (single detector only, \cite{GBM2009}), COMPTEL \citep{1993ApJS...86..657S}, \emph{Fermi}-LAT \citep{FermiLATPerformanceURL}, and \emph{Swift}-BAT \citep{2005SSRv..120..143B} (for photons with a 30\textdegree{} angle to boresight).}
\label{fig:effectiveAreaVsEnergy}
\end{figure}

\begin{figure}
\centering
\includegraphics[width=\columnwidth]{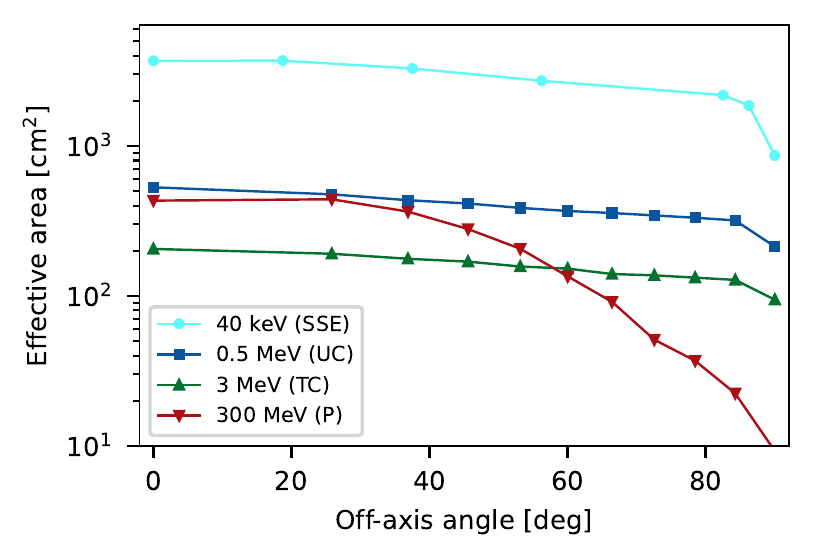}
\caption{Effective area as a function of the off-axis angle for various energies where each type of event dominates. The sudden decrease at a $\sim90^\circ$ angle is caused by the minimal projected area of the tracker layers as seen from the sides. Upwards-going events are severely attenuated by the calorimeter.}
\label{fig:effectiveAreaVsZenith}
\end{figure}

Figures \ref{fig:effectiveAreaVsEnergy} and \ref{fig:effectiveAreaVsZenith} show the effective area for the various types of events AMEGO-X will be able to detect (see Section \ref{sec:mission}), as a function of energy and the off-axis angle, respectively. The effective area for SSE peaks at about 40 keV with $\sim$3700 cm$^{2}$. This is about an order of magnitude greater than the effective area for UC events, the lowest energy events for which imaging is possible. In the 30-200 keV range, AMEGO-X's effective area is more than an order of magnitude greater than a single NaI detector of \emph{Fermi}-GBM. Although the \emph{Fermi}-GBM utilizes 12 such detectors, they have different pointings and therefore in all cases the total effective area always falls below that of AMEGO-X. This exemplifies why AMEGO-X will be able to detect fainter sources in this energy range compared to current detectors. 

\subsection{Background}
\label{sec:background}

We use MEGAlib to predict and simulate the average background AMEGO-X will observe during flight, including both instrumental and astrophysical components. The background simulations included prompt and delayed (activation-induced radioactive decays inside the instrument) emission from primary cosmic-ray electrons and positrons  \citep{Mizuno_2004}, protons and helium nuclei \citep{SPENVIS}, extra-galactic diffuse gamma-ray emission \citep{1999ApJ...520..124G}, and Albedo emission, i.e. secondary particles produced in the Earth's atmosphere, including neutrons \citep{Kole2015}, electrons and positrons \citep{200010, Mizuno_2004}, X-ray and gamma-ray photons \citep{tuerler2010, Mizuno_2004, 2009PhRvD..80l2004A, Harris2003}, and protons \citep{Mizuno_2004}. Cosmic-ray electrons and protons trapped in the Earth's geomagnetic field \citep{SPENVIS} only contribute via delayed emission, as their prompt emission is negligible outside the SAA, inside of which data-taking will be paused. The current AMEGO-X background model does not yet include the Galactic diffuse component. We assumed a 575 km orbit with an inclination of 6$^\circ$.

\begin{figure}[t]
\centering
\includegraphics[width=\columnwidth]{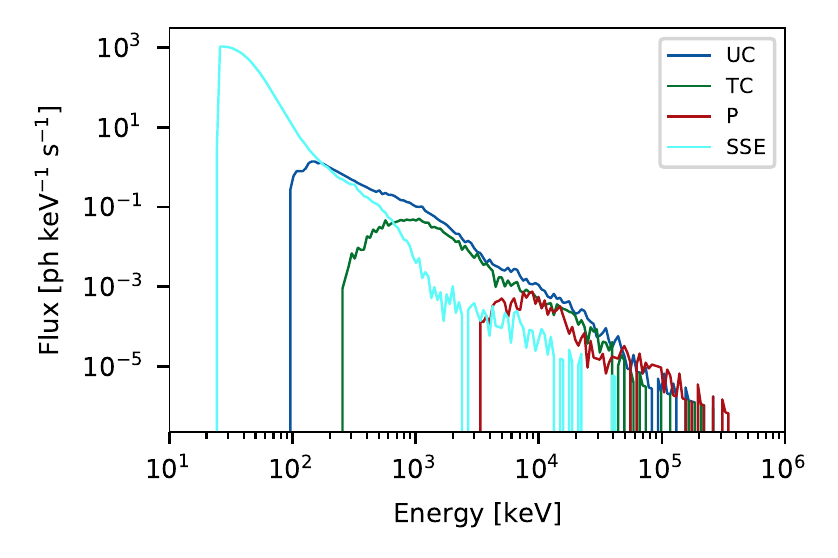}
\caption{AMEGO-X background rates including both astrophysical and instrumental contributions. The background is comprised of (from left to right) single-site events (cyan), untracked Compton events (blue), tracked Compton events (green), and pair events (red).} 
\label{fig:BG}
\end{figure}

Figure~\ref{fig:BG} shows the expected measured background spectra for the four different event types. The statistical fluctuations are due to the short simulation time. In general, calculation of the SSE background is very computationally expensive, and so we use a simulation time of 60 seconds for this event type (a long enough exposure to cover short and long GRBs) and 1 hr for the rest. For SSE the background counts rate between 25 - 100 keV is $\sim$ 23 kHz.  For comparison, the background counts rate for UC events in the energy range 25 - 100 keV (100 keV - 1 MeV) is $\sim$ 2 Hz ($\sim$ 330 Hz).  

%%%%%%%%%%%%%%%%%%%%%%%%%%%%%%%%%
%Note that the BG has been scaled by 0.8 
%to account for the smaller pixel area.  
%%%%%%%%%%%%%%%%%%%%%%%%%%%%%%%%%
\subsection{Sensitivity}

In order to estimate the sensitivity we define the signal-to-noise ratio (SN) as 
\begin{equation}
    SN = \frac{S}{\sqrt{S+B}} \ ,
\end{equation}
where $S$ is the number of signal counts and $B$ is the number of background counts. The SN is computed for each event type, and then sum in quadrature if needed. We chose a conservative threshold for detection  $SN > 6.5$ corresponding to a false alarm rate of $\sim1$ yr$^{-1}$. For a point-like source, we have $S=F\cdot A \cdot T$ and $B= b \cdot T$ , where $F$ is the energy-integrated gamma-ray particle flux, averaged over the observing time $T$, $A$ is the spectrum-averaged effective area determined from simulations, and $b$ is the background rate, also determined from simulation. $A$ and $b$ depend on the event type in question (SSE or UC). For a given signal-to-noise ratio $SN$ and event type $x$, and fixed spectral shape, we can solve for $F$ to determine the minimum flux $F_{SN}^{x}$ that would produce a detection.

We estimate the sensitivity for a fiducial 1 s burst at 30$^{\circ}$ off-axis as $F_{6.5}^{SSE}$ $\sim0.5 \ \mathrm{ph \ cm^{-2} \ s^{-1}}$ using only SSE between 25 keV and 100 keV. In Section \ref{sec:science} we discuss in detail the sensitivity for various realistic models of GRBs and magnetars. 

\begin{figure}[t]
\centering
\includegraphics[width=0.5\textwidth]{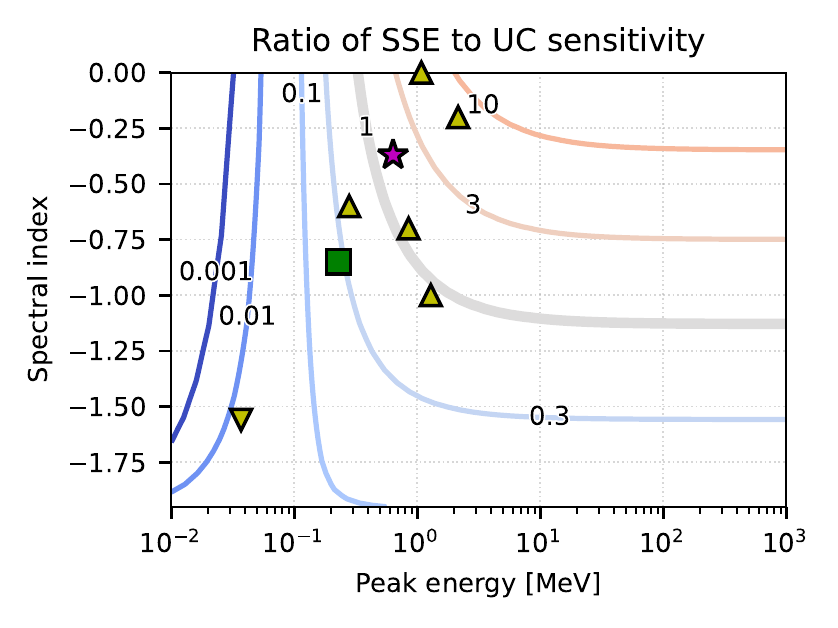}
\caption{Ratio of the integral flux sensitivity (minimum flux between 10 keV and 10 MeV needed for a 6.5$\sigma$ detection in 1 s) achieved by SSE alone to the sensitivity achieved in UC events alone, as a function of the spectral parameters: peak energy $E_{peak}$ and low-energy spectral index $\gamma$. We assume a Comptonized spectral model, described in the text. A ratio of $\leq$1 implies that the addition of SSE significantly enhances the sensitivity of the instrument. SSE dominate the sensitivity for peak energies below hundreds of keV and for soft spectral indices. The purple star represents the ``nominal'' population of short GRBs considered in Section \ref{sec:grbs}, while the green square shows the case of GRB 170817A. The yellow upward pointing triangles mark the main peak emission of magnetar giant flares (discussed in Sec.~\ref{sec:mgf}), while the downward pointing triangle represents the SGR1935+2154 burst associated to an FRB reported in 2020 \citep{2020ApJ...898L..29M,2021NatAs...5..378L}.} 
\label{fig:sensi_ratio_2D}
\end{figure}

The gain in sensitivity by including SSE events in a given analysis depends on the spectral characteristics of the source. Figure \ref{fig:sensi_ratio_2D} shows the ratio $F_{6.5}^{SSE}/F_{6.5}^{UC}$ between SSE and UC sensitivity for bursts with characteristic Comptonized energy spectra modeled as 
\begin{equation}
   \frac{dN}{dE} \propto \left( \frac{E}{E_0}\right)^{\gamma}  \exp\left( - \frac{E \left( \gamma + 2 \right)}{E_{peak}}\right), \,
 \label{eq:Comptonize_spec}
\end{equation}
where $E_{peak}$ is the peak energy, $\gamma$ is the low-energy spectral index and $E_0$ is the arbitrary pivot energy. In general, the analysis of sources with either $E_{peak} \lesssim 300$ keV or $\gamma \lesssim -1.15$ benefit significantly from the inclusion of SSE, as well as some other combinations.  Observe that the Comptonized form in Eq.~\ref{eq:Comptonize_spec} is well suited to describing many GRB and magnetar burst spectra.

\subsection{Localization}

AMEGO-X is capable of reconstructing the direction of a single photon if its passage through the instrument leaves at least two energy deposits $>$25 keV, whether on two or more tracker layers, or on a single tracker layer and the calorimeter. These photons typically have an energy $>$100 keV and interact due to either Compton scattering or pair production. Using this information AMEGO-X will localize a transient source (5 $\mathrm{ph \ cm^{-2} \ s^{-1}}$ between 120 keV and 1 MeV, 1 s in duration) within a $<2^\circ$ radius (90\% cont).

Single-site events, on the other hand, are primarily due to photoelectric interactions and by definition lack any track. The only information recorded is their deposited energy, time, and location in the tracker. While an event-by-event direction reconstruction is therefore impossible, we can use the aggregate information of multiple SSE to estimate the sky location of the source that produced them.

\begin{figure*}
\centering
\gridline{\fig{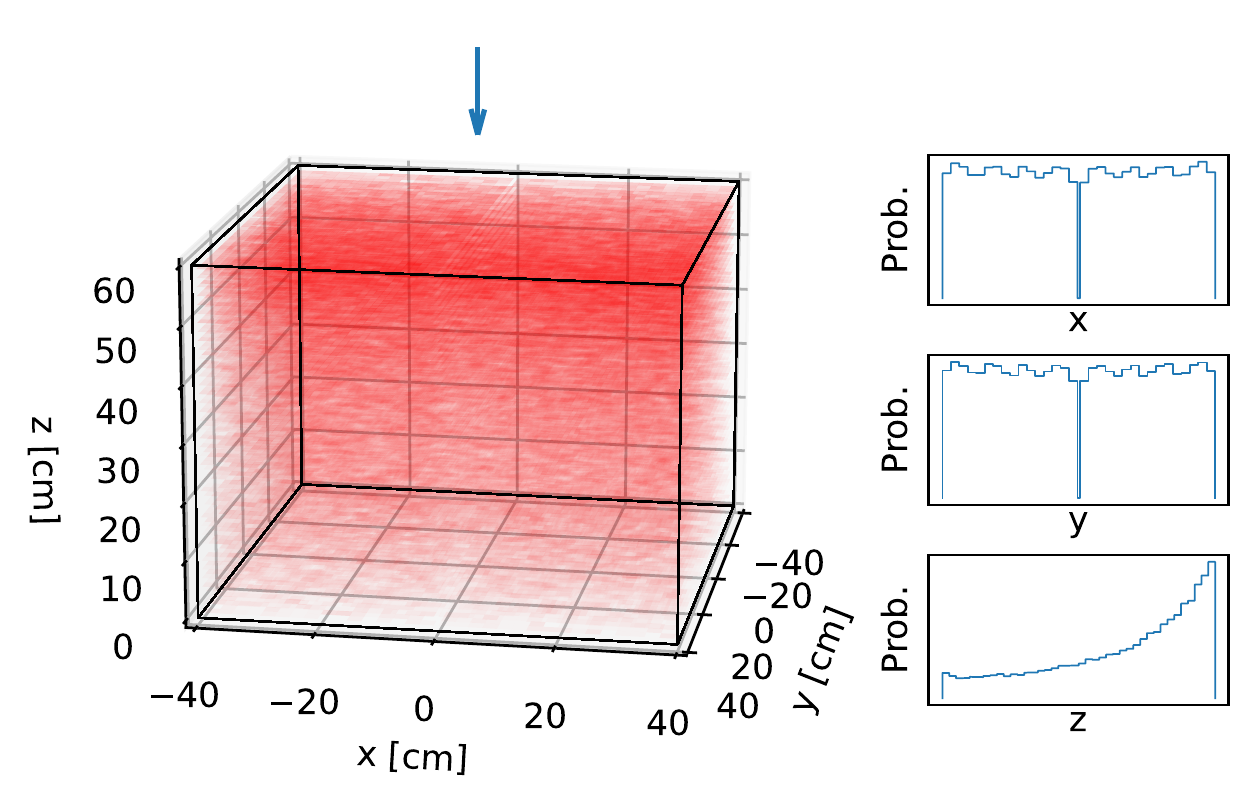}{0.48\textwidth}{(a)}
          \fig{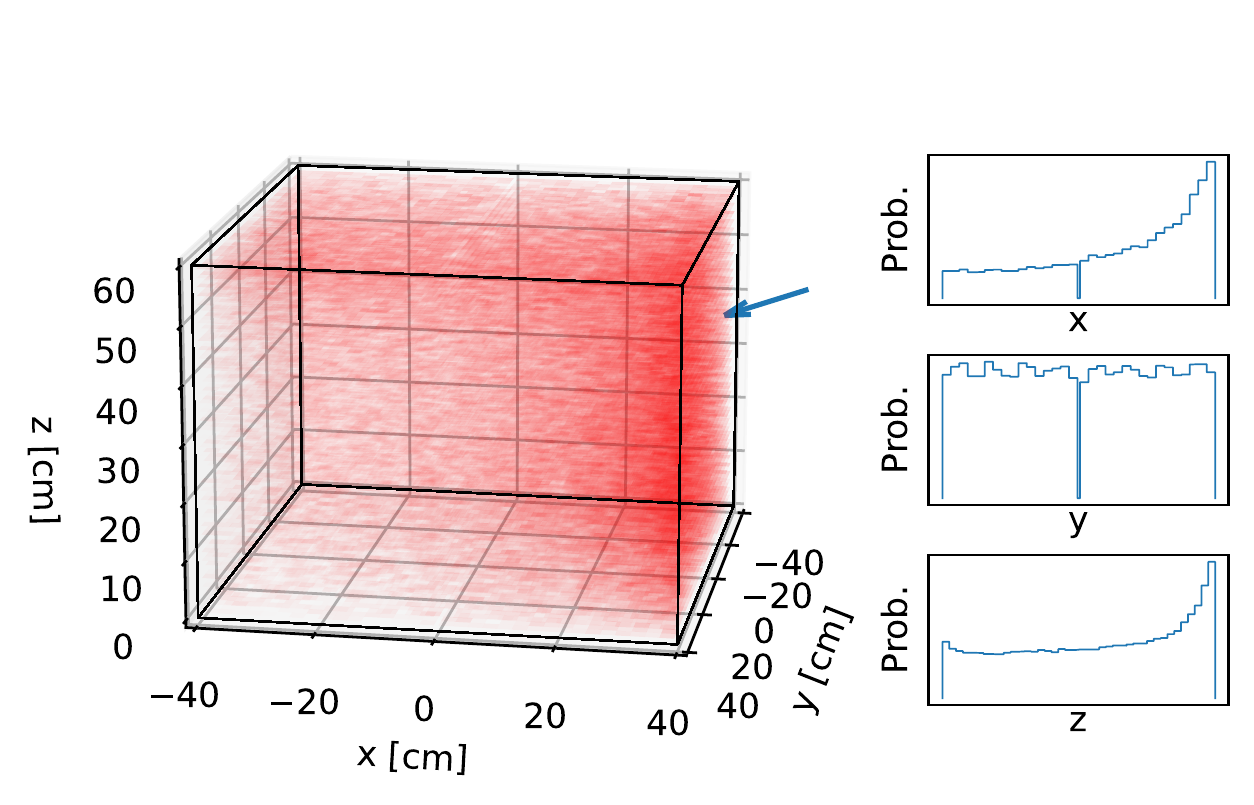}{0.48\textwidth}{(b)}}
\caption{Expected SSE position probability distribution from an on-axis burst (a) and one located 70$^\circ$ off-axis and along the $xz$-plane (indicated by the arrows). The insets show the projection along each axis.  These distributions vary as a function of the incoming direction due to the partial attenuation by the different tracker layers and provide the leverage to estimate the source location. The null count at $x=0$ and $y=0$ is caused by the spacing between the four towers that compose the tracker.}
\label{fig:locdist}
\end{figure*}

The method is similar to those employed by other count-based detectors, such as BATSE \citep{Briggs1999-BATSELoc} and \emph{Fermi}-GBM \citep{Connaughton2015-GBMLoc}, and relies on the change in relative acceptance of each detector as a function of the incoming direction. BATSE and GBM relied primarily on the different pointing of each detector --- with an approximately a cosine angular response. These missions also take into account the scattering and attenuation by the different spacecraft structures, although it was not the dominant source of leverage when fitting the source location. The opposite is true for AMEGO-X. Even though all layers in the tracker are pointing towards the same direction, their mutual shadowing leads to abundant directional information, as shown in Figure \ref{fig:locdist}. A similar principle was followed by POLAR \citep{WANG2021164866}.

The localization sky maps are obtained through a maximum likelihood analysis. It compares the number of detected events at various spatial coordinates in the tracker to the expectation from a source located at a given sky coordinate, and finds the best match. We construct the test statistic ---called a \textit{C}-stat estimator in other contexts \citep{1979ApJ...228..939C}---

\begin{eqnarray}
TS &= 2 \frac{\displaystyle\max \left(\sum_i \log P(d_i; f) \right)}{\displaystyle \sum_i \log P(d_i; f = 0)} \nonumber\\
P(d_i; f) &= \frac{\displaystyle (b_i + e_i f)^{d_i} e^{-(b_i + e_i f)}}{\displaystyle d_i!} ,
\end{eqnarray}

where

\begin{itemize}
\item $P(d_i; f)$ is the Poisson probability of observing $d_i$ events given a source with a flux $f$;
\item $b_i$ is the estimated number of background events;
\item $e_i$ is the expected excess given a source spectral hypothesis, per flux unit. It is a function of the sky coordinate of the hypothetical source.
\end{itemize}

The index $i$ runs through all the different tracker locations where a hit can be found in the tracker. Each pixel of each tracker layer can be considered as a individual detector. Although the tracker pixel size is 1 mm, it was sufficient to divide each of the 40 tracker layers (80cm$\times$80cm) into a 32x32 grid, uniformly spaced along each dimension (as in Figure \ref{fig:locdist}).

The unitary expected excess values $e_i$ are obtained from look-up tables previously filled using simulations. The $i$-th element of each table contains the differential effective area for events arriving from a given sky location. The differential effective area equals the total effective area of the instrument multiplied by the fraction of the events detected in the $i$-th tracker location. We use a HEALPix sky pixelization \citep{healpix2005} to describe the incoming direction, although other schemes are also possible.

\begin{figure}
\centering
\includegraphics[width=.5\textwidth]{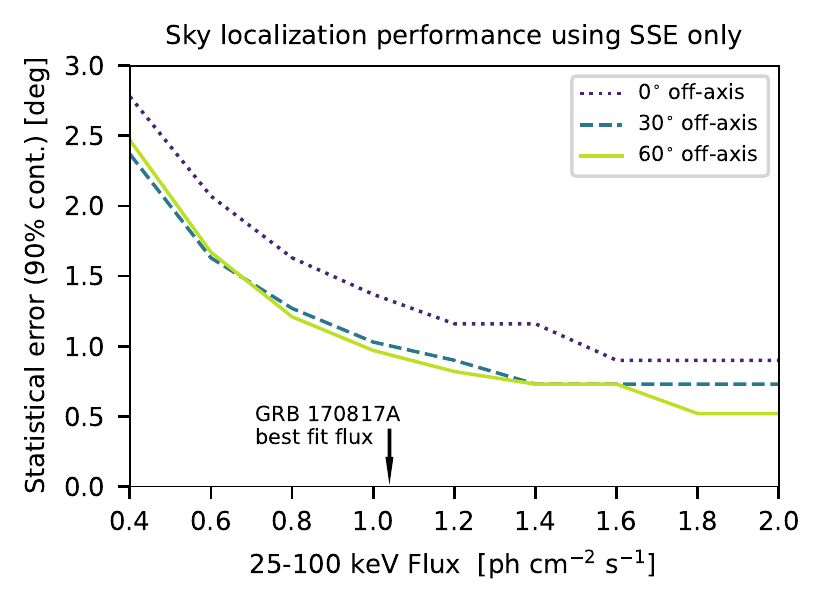}
\caption{Expected localization uncertainty from SSE only for a fiducial 1 s burst as a function of the flux between 25 keV and 100 keV, and the off-axis angle. While the localization error region is not well-described by a disc in general, this represents the radius containing the equivalent solid angle area. As described in the text, systematic errors will need to be considered in addition to the statistical-only uncertainty presented here. Compton and pair events can further reduce the localization error.}
\label{fig:loc_uncertainty}
\end{figure}

The $TS$ value is computed for each sky location ---the flux is considered a nuisance parameter for this purpose. A sky location confidence interval is then obtained based on Wilk's theorem \citep{10.1214/aoms/1177732360} and considering two degrees of freedom ---i.e. a 90\% containment corresponds to $\Delta$TS$<4.60$ with respect to the maximum, which we confirmed for our case through simulations. This calculation was done for simulated sources injected at various sky locations, using the estimated background presented in Section \ref{sec:performance}. The expected performance is shown in Figure \ref{fig:loc_uncertainty}. 

In addition to the expected statistical errors showed in Figure \ref{fig:loc_uncertainty}, the main source of systematics that can contribute to the localization unertainty consists of the inaccuracies of the look-up tables used to obtain the expected excess counts as a function of the incoming direction. This can be mitigated by a meticulous calibration campaign on ground. Other sources of systematics include the mismodeling of the source spectrum using a parametric shape, as well as possible errors simulating the effect of photons scattered by the Earth's atmosphere \citep{Connaughton2015-GBMLoc, palit2021revisiting}. During flight, these systematic errors can be studied and mitigated by performing both a cross-calibration with other instruments, and a self-calibration with higher-energy photons that do produce tracks in the detector.
\section{Science enabled by utilizing single-site events}
\label{sec:science}

\subsection{Gamma-Ray Bursts}
\label{sec:grbs}

Gamma-Ray Bursts (GRBs) are a class of transient events characterized by a highly variable pulse-like signal in gamma-rays, known as the prompt phase, followed by a smoothly decaying emission at various wavelengths, called the afterglow. The prompt emission can last from $<2$ s in the case of bursts classified as short GRBs (SGRBs), up to hundreds of seconds for long GRBs (LGRBs). It is generally believed that most LGRBS result from the core-collapse supernovae, while the progenitors of SGRBs are compact object mergers, such as binary neutrons stars (BNS) and neutron star - black hole binaries \citep{levan2016gamma}.   

In recent years short GRBs have played a key role in the development of multi-messenger astronomy after the simultaneous detection of GRB170817a and the gravitational wave (GW) event GW 170817 \citep{grbgw170817}, confirming BNS mergers as a type of progenitor.  Questions still remain about the physics of neutron stars, jet formation and structure, acceleration mechanisms and the existence of other progenitor types. In order to answer them we need a higher number of such joint GRB-GW detections, followed by extensive  observations of the afterglow. 

AMEGO-X can make use of its large field of view and on-board Transient Alert system to send rapid alerts with precise localization information to other detectors. Since the emission of GRBs typically peaks at hundreds of keV, with a median energy peak of $\sim$200 keV \citep{gbm_catalog_spectra_2014}, it is expected that SSE will play a major role in the detection and analysis of the majority of this type of transient phenomena by AMEGO-X. 

In order to investigate this we first calculate the sensitivity using SSE alone. We use a fiducial model for SGRBs based on the average properties of the GRBs detected by GBM~\citep{2020ApJ...893...46V}. The model consists of a Comptonized spectrum ---described in Section \ref{sec:performance}--- with $\gamma = -0.37$ and $E_{peak}=$ 636 keV.  We use a flat lightcurve lasting 1 s.  The choice of a Comptonized spectral form with two free parameters is made for simplicity.  Yet we note that it is preferred over the well-known Band model \citep[a modified broken power law with 4 parameters:][]{Band-1993-ApJ} for many of the bursts in the most recent {\it Fermi}-GBM spectral catalog in \cite{Poolakkil-2021-ApJ}.  Thus the Comptonized form is suitable for developing a general sense of the utility of SSE in enhancing the AMEGO-X low energy sensitivity.

\begin{figure}[t]
\centering
\includegraphics[width=\columnwidth]{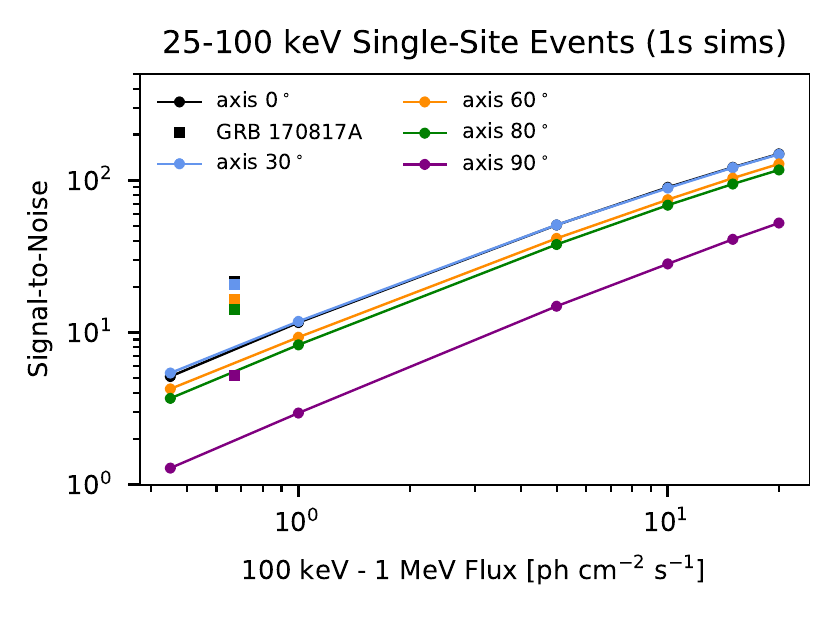}
\caption{Signal-to-noise ratio (defined in text), using SSE only, versus flux for the fiducial SGRB model described in text. We include counts between 25 - 100 keV. The curves are for the representative models, and the squares are for GRB170817A. The source is simulated at the different off-axis angles. Note that the on-axis case is mostly behind the 30$^\circ$ off-axis case. } 
\label{fig:SGRB_SNR}
\end{figure}

\begin{figure}
\centering
\includegraphics[width=\columnwidth]{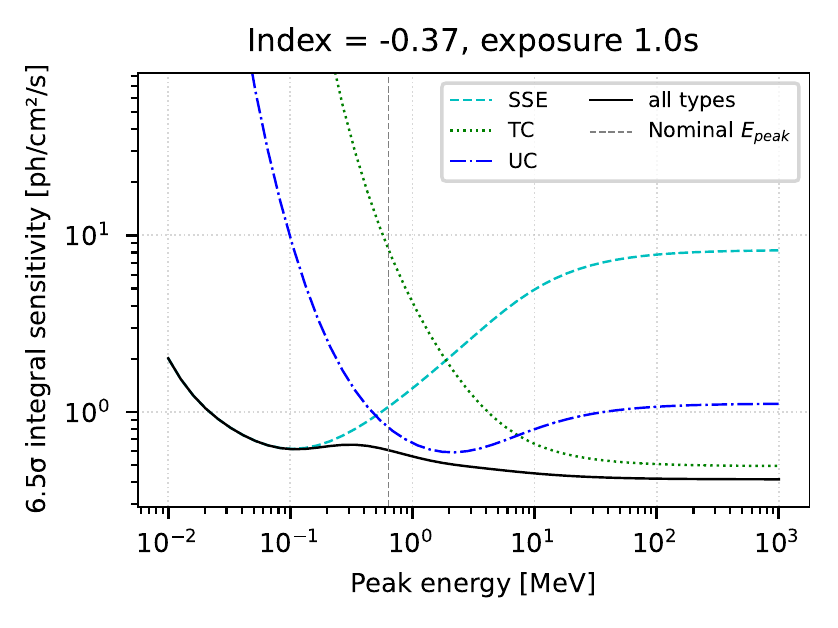}
\caption{Predicted sensitivity between 10 keV and 10 MeV as a function of the peak energy for a Comptonized spectrum with a fiducial low-energy spectral index $\gamma = -0.37$ and 1 s duration. Single-site events  (SSE) dominate the sensitivity for peak energies below hundreds of keV over tracked (TC) and untracked (UC)  Compton events.} 
\label{fig:SGRB_Sensi}
\end{figure}

As shown in Figure \ref{fig:SGRB_SNR}, we vary the flux and find where $SN=6.5$, defining our detection threshold. The estimated sensitivity is $6.5 \times 10^{-1} \ \mathrm{ph \ cm^{-2} \ s^{-1}}$ between 100 keV and 1 MeV (corresponding to $5.2 \times 10^{-1} \ \mathrm{ph \ cm^{-2} \ s^{-1}}$ between 50 keV and 300 keV). We also simulated GRB170817A using the best fit parameters $\gamma =-0.85$ and $E_{peak}=$ 229 keV~\citep{2017ApJ...848L..14G}. A similar event will be detected with high SN for the majority of the field of view. Figure \ref{fig:SGRB_Sensi} also puts the SSE sensitivity of GRBs in perspective with the Compton-only sensitivity by taking the SGRB fiducial model and varying the peak energy. 

%%%%%%%%%%%%%%%%%%%%%%%%%%%%%%%%%

%%%%%%%%%%%%%%%%%%%%%%%%%%%%%%%%%

The increase in sensitivity for peak energies below hundreds of keV will be reflected in the number of detections. We quantify this by using empirical distributions of GRB peak energies, spectral indices and flux values based on \textit{Fermi}-GBM observations \citep{2020ApJ...893...46V}, extended below the threshold based on the Swift-BAT observations \citep{2016ApJ...829....7L}. We assumed that the observed logN-logS curve reflects the intrinsic GRB flux distribution, and considered a power law with an exponential cutoff spectrum. Based on these assumptions, we synthesized more than $10^5$ events, convolved them with the response of the instrument, and computed the SN for each simulated event for both single-site and Compton events. As opposed to BATSE and Fermi-GBM, which use scintillators, AMEGO-X does not require a minimum of two detectors to trigger in order to avoid false detections, mostly from phosphorescence events. This allows us to use the full effective area of the detector to trigger on-board.

\begin{figure}[tbp]
\centering
\includegraphics[width=\columnwidth]{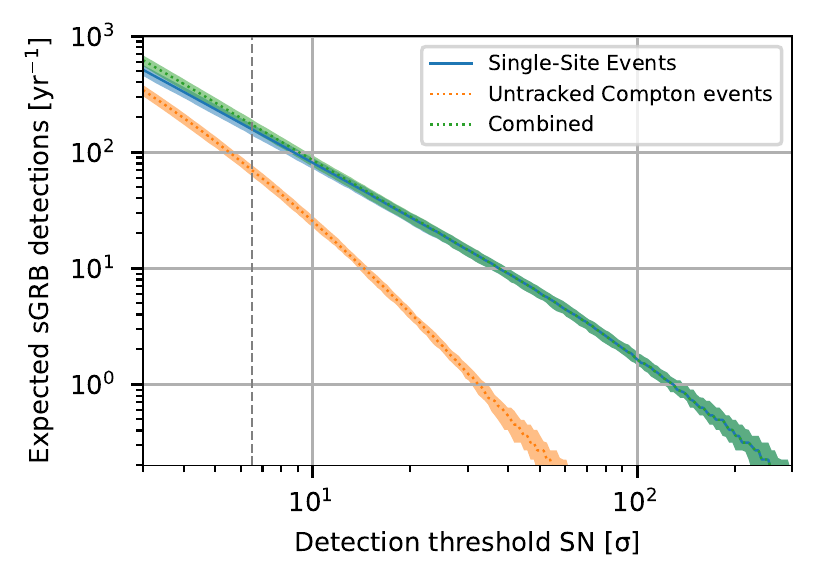}
\caption{Expected rates of short GRBs seen by \mbox{AMEGO-X} above various detection thresholds. The filled bands correspond to the theoretical uncertainty in the modeling. Including single-site events significantly increases the rate of short GRB detections (nominal threshold 6.5$\sigma$ shown by the dashed line) as well as the rates of short GRBs detected with high enough significance to enable precise localization and detailed studies of the spectrum and lightcurve.} 
\label{fig:SGRB_rates}
\end{figure}

We find that by utilizing SSE, AMEGO-X will be able to detect $\sim$200 $yr^{-1}$ short GRBs, more than doubling the number of detections compared to using only Compton events. As shown in Figure \ref{fig:SGRB_rates} not only will the overall number of detections will increase, but the use of SSE will improve the signal-to-noise ratio overall allowing for more detailed spectral and time-dependent analyses. 

%%%%%%%%%%%%%%%%%%%%%%%%%%%%%%%%%
%\begin{figure}[tbp]
%\centering
%\includegraphics[width=0.8\textwidth]{figures/longGRBrates_log.pdf}
%\caption{Expected rates of long GRBs seen by AMEGO-X above various detection thresholds. The filled bands correspond to the theoretical uncertainty in the modeling. Including single-site events significantly increases the rate of long GRB detections (nominal threshold 6.5$\sigma$ shown by the dashed line) as well as the rates of long GRBs detected with high enough significance to enable precise localization and detailed studies of the spectrum and lightcurve.} 
%\label{fig:LGRB_rates}
%\end{figure}
%%%%%%%%%%%%%%%%%%%%%%%%%%%%%%%%%

\subsection{Magnetar giant flares and short bursts}
\label{sec:mgf}

Magnetars are highly magnetized ($>10^{13}$ G), young neutron stars (NSs) that display a wide range of radiative activity \citep{2008A&ARv..15..225M,Turolla2015}. Among the different dramatic variability exhibitions, magnetar giant flares (MGFs) are the most energetic. They are characterized by a short ($\lesssim 100$ ms), bright ($10^{44}-10^{46}$ erg), hard spike immediately followed by a $10^{44}$ erg/s minutes-long quasi-thermal ($3 k_b T \sim 50-100$ keV) pulsating tail modulated by the spin-period of the magnetar: these signatures have been observed for all three nearby events  reported in ~\cite{Mazets1979}, \cite{Mazets1999} and \cite{Hurley2005,2005Natur.434.1107P}.

In distant events only the prompt spike is currently detectable by wide-field GRB instruments and thus MGFs resemble cosmological short GRBs. As opposed to Galactic MGFs, which saturate all viewing detectors during the prompt emission, characterization is possible for extragalactic ones. The recently determined intrinsic volumetric rate of MGFs places MGFs as the dominant channel for extragalactic gamma-ray transients \citep{Burns2021}. Moreover, the 2020 MGF from the Sculptor galaxy \citep{2021NatureSvinkin, Roberts2021} revealed a peak energy of $\sim$1.5 MeV and a spectral index $\sim$0, resulting in a harder spectrum compared to typical cosmological short GRBs: this would explain the small number (only 4 in ~27 years) of extragalactic MGFs detected so far by GRB monitors, which have trigger systems optimized to detect the latter kind of short GRBs \citep[see Figure 3 of ][]{Negro2021S3}.  Given these spectral characteristics, the majority of the AMEGO-X sensitivity to MGFs comes from UC events, as shown in Figure \ref{fig:sensi_ratio_2D} (yellow upward pointing triangles). 

Considering the intrinsic rate of MGFs found by \cite{Burns2021}, we estimate that AMEGO-X will detect 1 to 19 local MGFs within 25 Mpc for a 3 year mission. The detection of just six MGFs, either Galactic or extragalactic, would double the current statistics of the known population \citep[we exclude the first MGF from the LMC in 1979 which was not accounted for in the intrinsic rate computation of][]{Burns2021}. This increase would improve the current statistical uncertainty on the average index of the power law spectral fits for MGF initial spikes, thereby better securing the intrinsic energetics distribution for MGFs, and consequently reducing the volumetric rate uncertainty of MGFs by $\sim$40\%. Such an improvement is crucial to determining the favoured formation channel: only progenitors ---e.g. core-collapse supernovae--- with comparable or higher rates could form magnetars. Furthermore, a more stringent constraining of the cosmic MGF rate would help determine whether or not there is more than one MGF per magnetar during its active lifetime: this could then help identify the first class of repeating short GRBs ever observed.

\begin{figure*}
\centering
\gridline{\fig{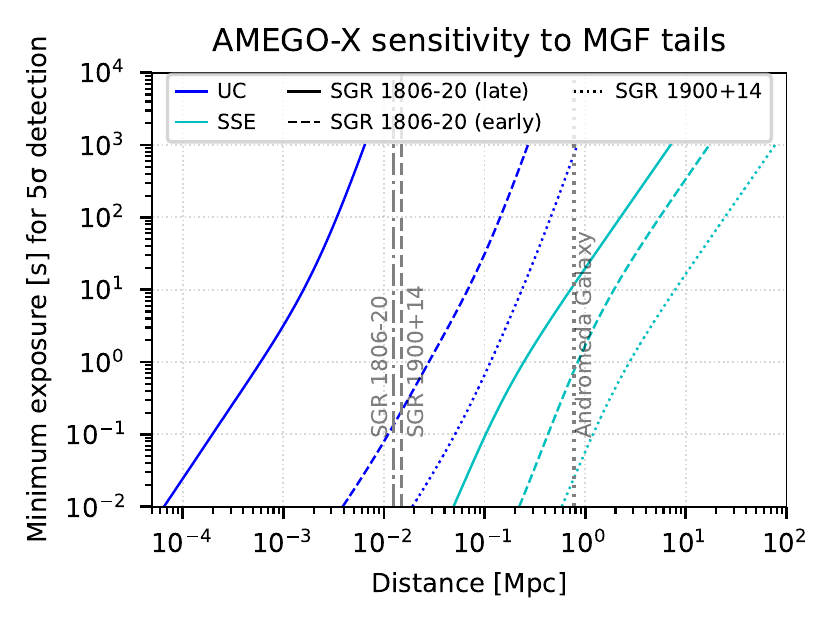}{0.48\textwidth}{(a)}
          \fig{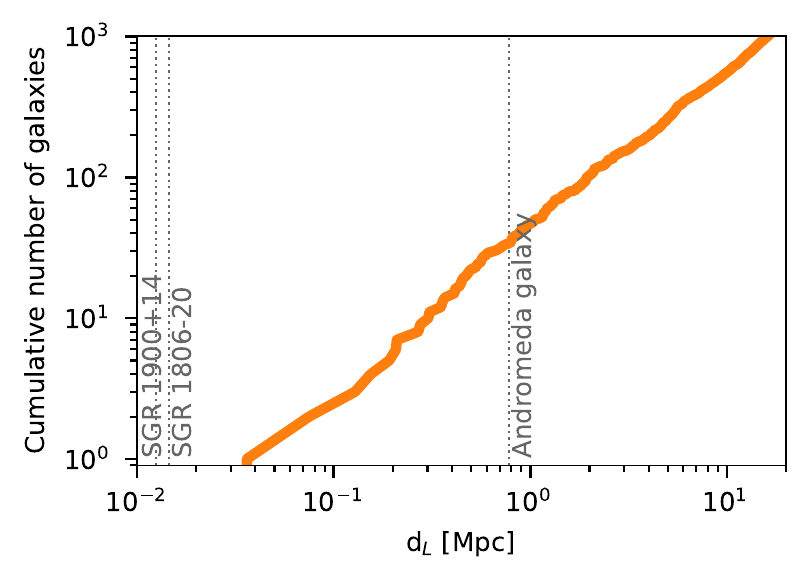}{0.46\textwidth}{(b)}}
\caption{(a) Minimum exposure time necessary for AMEGO-X to detect average MGF tail emission as a function of the distance to the magnetar, using SSE (teal) and Compton only (blue). Results are shown for three different spectral models: A surface blackbody spectrum with kT = 5.1 keV (solid lines) representative of the late-tail emission of the MGF from SGR 1806-20 between $\sim$272 and 400 seconds, a blackbody spectrum with kT = 9 keV (dashed line) representative of the first 200 seconds of the tail emission from SGR 1806-20, and an OTTB+PL model spectrum with kT = 25 keV, and $\Gamma = 1.47$ (dotted lines) representative of the tail emission from SGR 1900+14. The spectra of the MGF tail from SGR 1806-20 are from \cite{Hurley2005}, while the spectrum for the MGF from SGR 1900+14 is from \cite{Feroci1999}. The grey vertical lines indicate the measured distances to the two magnetars and to the Andromeda galaxy. With SSE, AMEGO-X would be able to detect emission from magnetar tails similar to the ones already observed on second timescales even if they were located in the Andromeda galaxy or other nearby galaxies. With Compton events only, AMEGO-X could still expect to detect tail emission from some galactic MGFs. (b) Cumulative distribution of nearby galaxies.}
\label{fig:sensi_magnetarTail}
\end{figure*}

For nearby MGFs with a few Mpc, the observation and characterization of MeV-GeV afterglows of MGFs enables the determination of the outflow/jet bulk Lorentz factor via pair creation transparency arguments \citep{Roberts2021}.  Assuming an event like the 2004 MGF from SGR1806-20 (GRB 041227), we estimate the maximum distance AMEGO-X would have detected the tail. We rely on the spectral and time evolution study reported in \cite{hurley2005exceptionally}, which gives a blackbody (BB) spectral shape with temperatures decreasing with time between $\sim$9 keV and $\sim$4 keV (see Figure 1(b) of that paper). In Figure \ref{fig:sensi_magnetarTail} we show, for both SSE and UC events, the maximum distance for a given exposure time that AMEGO-X would have detected the tail of GRB 041227 at 5$\sigma$ significance: the advantage of SSE is clear. For the brightest part of the tail (between 30 and 200 seconds after the main peak), considering an exposure of 0.5 seconds, which would be enough to detect the periodicity due to the star rotation, AMEGO-X would detect the emission from the tail at a distance of $\sim$700 kpc with SSE; with UC events only, this distance drops to 20 kpc, which is about the distance of SGR 1806-20.  During the fainter, latter half of the tail, after 270 s after the initial spike, the temperature drops to an average of 5.1 keV, and SSE would contribute to observations of similar tail portions only out to a distance of about 160 kpc.  With an integration time of 10 seconds, time for which we can assume a stable background, AMEGO-X would recover the tail of a MGF from a magnetar in the Andromeda Galaxy (about 765 kpc away), which would be a first.

The blackbody spectral fit in \cite{hurley2005exceptionally} for the GRB 041227 tail, however, significantly underestimates the measured spectrum above 30 keV, where an additional power-law (PL) component would better represent the data. This additional component would further improve the AMEGO-X SSE sensitivities, resulting in an extension of the distance range for detectability of MGF tails. 
Such additional PL component was even more evident in the MGF tail from SGR 1900+14 in 1998 \citep{Feroci1999}: the BB component was peaking at higher temperatures and the additional PL component was dominant above 100-200 keV with a spectral index steepening with time, and disappearing after about 200 seconds. Assuming the spectrum measured in the first 67 seconds of the events by \cite{Feroci1999}, namely an optically thin thermal bremsstrahlung (OTTB) component with kT$\sim$25 keV, plus a power law with index $\sim$1.5, the 5$\sigma$ detectability as function of the source distance and exposure time is reported in the left panel of  Figure \ref{fig:sensi_magnetarTail} (dotted lines).  AMEGO-X SSE events would allow the detection of such a tail emission out to $\sim 2$ Mpc  with an exposure of 0.5 seconds, and with an exposure of 10 seconds it would detect it out to $\sim 8$ Mpc. In the right panel of Figure \ref{fig:sensi_magnetarTail}, for reference, the cumulative distribution of galaxies is presented as a function of the distance from Earth. Clearly, the enhancement of AMEGO-X sensitivity enables the extension of observations of MGF tails to more host galaxies.  MGFs tails may also appear unassociated with the initial hard spike in the particular case that the latter is due to a collimated outflow \citep{Roberts2021} directed outside our field of view.  Accordingly, population studies of potential “orphan” tails enables independent constraints for outflow collimation, thereby significantly enhancing the SSE contribution to AMEGO-X extragalactic MGF science.

%All three Galactic MGF also exhibited quasi-periodic oscillations (or QPOs) in the ensuing tails \citep{2008A&ARv..15..225M}. Likely associated with oscillations crust+core caused by the giant flare, such eigenmodes could excite kHz neutron star f-modes which damp via GWs. These will be accessible in the nearby universe by proposed third generation GW detectors \cite{Macquet2021}. Characterization of these modes in EM and GW sectors also enables asteroseismology and understanding of the NS equation of state. Various QPOs were observed in, for example, the tail emission of SGR 1806–20 with the most significant one at 92.5 Hz \citep{SGR1806-20QPO}. In Fig.~\ref{fig:sensi_magnetarTail} we show the sensitivities down to exposure times of 0.01 seconds (100 Hz frequency): for such short exposure times the maximum detection distance for SSE is about 200 kpc for a MGF like the one from SGR 1806-20, and 600 kpc for a MGF like the one from SGR 1900+14. This means that AMEGO-X would detect QPOs with frequency below 100 Hz out to greater distances then the ones quoted above (higher frequencies QPOs could be detected at closer distances as can be deduced in Fig.~\ref{fig:sensi_magnetarTail}).

In addition to the MGFs, magnetars also produce much more numerous short duration (0.05-0.5 s) bursts with much lower luminosity (peak luminosity ranging between $10^{36}-10^{43}$ erg/s, with a power-law event energy distribution). In April 2020, the magnetar SGR 1935+2154 entered an active state, producing a ``burst storm" with more than 200 bursts in about a thousand seconds (see e.g. \cite{Younes2020_Storm}). A few hours after the peak of the storm, a hard X-ray burst (FRB-X) was observed contemporaneously with a fast radio burst \citep[FRB,][]{2020ApJ...898L..29M,2021NatAs...5..378L,Bochenek_2020,2020Natur.587...54C}.  \cite{2021NatAs...5..378L} reported the time integrated ($T_0$-$0.2s-T_0$+$1s$) spectrum of the FRB-associated burst to be well represented by a soft cutoff-power-law (CPL) with photon index of 1.5 and energy cutoff around 84 keV (corresponding to a peak energy of 37 keV). This spectrum is somewhat steeper and extends to high energies relative to spectra of other bursts in the same epoch \citep{Younes2020_FRBSGR}.  Note that despite the temporal association, it is unclear whether FRB-X and the FRB are causally connected.

Figure \ref{fig:sensi_ratio_2D} highlights the clear benefit SSE will bring to the AMEGO-X mission for such a soft, fast (1 s integration time) transient. Assuming the best-fit CPL model reported in \cite{2021NatAs...5..378L}, the energy-integrated flux between 10 and 100 keV is about 77.5 ph/cm$^2$/s, while AMEGO-X minimum flux detectable at 6.5$\sigma$ (considering 1 s exposure) is 50 ph/cm$^2$/s for UT events and $\sim$0.5 ph/cm$^2$/s forSSE. The latter consideration means that AMEGO-X could detect bursts which are a factor $\sim 100$ dimmer than the FRB-associated burst from SGR 1935+2154. This is particularly interesting since several additional, dimmer FRBs have been observed from SGR 1935+2154 \citep{moreFRBnoSGR, FASTfrb}. Identifying and collecting a much larger sample of magnetar short bursts afforded by sensitive wide-field monitoring capability of AMEGO-X SSEs, in concert with wide-field radio facilities in the 2020s, will help solve the mystery of why not all magnetar short bursts produce FRBs. Moreover, millisecond time-offsets between the radio and hard X-rays, as observed in SGR 1935+2154 by INTEGRAL \citep{2020ApJ...898L..29M}, will also help inform on the burst mechanism.

Magnetar bursts have a rate of about a few thousand per decade in \textit{Fermi}-GBM \citep{2015ApJS..218...11C,2020ApJ...902L..43L}, but are highly clustered in time in specific sources and exhibit statistical behavior and unpredictability similar to earthquakes \citep{1996Natur.382..518C,1999ApJ...526L..93G}. The short bursts are typically characterized by spectral fits employing double blackbodies of similar luminosities, with a hot component at $k T_h \sim 20-50$ keV and a cooler component at $k T_c\sim 5-10$ keV \citep{2008ApJ...685.1114I,2012ApJ...749..122V,2014ApJ...785...52Y,2020ApJ...893..156L}. Thus, magnetar short bursts are expected to be readily observable in the SSE channel with AMEGO-X, with a higher rate than GBM. Short bursts have also exhibited quasi-periodic oscillations (QPOs) \cite[e.g.,][]{2014ApJ...787..128H} that will be important to identify in SSEs.

\section{Summary and conclusions}

In this paper it has been shown that the performance of AMEGO-X at low energies is significantly improved by utilizing photons that deposit all their energy in a single tracker pixel, further advancing its science goals. Using single-site events (SSE) we lower the AMEGO-X detection energy threshold and achieve a continuous sensitivity to transient events over five orders of magnitude, from 25 keV to $\sim$1 GeV.

While SSE events leave no tracks, we have demonstrated that we can use the aggregate signal from a transient source to provide a competitive sky localization and support the multi-wavelength and multi-messenger efforts of the community. An event similar to GRB170817A would be localized to within a $<$2$^\circ$ radius. 

We find an integrated flux sensitivity (6.5$\sigma$) between 25 keV and 100 keV of $\sim0.5 \  \mathrm{ph \ cm^{-2} \ s^{-1}}$. Employing SSE gives a significant benefit for studying transient sources with peak energies below a few hundred keV or with an otherwise soft spectrum. It more than doubles the number of detected gamma-ray bursts that AMEGO-X will detect and allows a better understanding of the source spectral and time-dependent properties.

SSE capability will open the possibility for many interesting studies relating to magnetar giant flares (MGFs). The tails of MGFs can be detected up to a distance of $\sim 700$ kpc with a time resolution of 0.5 s, enough to observe the periodicity due to the star rotation. Integrating for $\sim 10s$ AMEGO-X could detect a bright MGF tail like the one in 2004 from SGR 1806-20 as far as the Andromeda Galaxy. The achieved sensitivity also enables the detection of soft gamma-ray bursts associated with FRBs that are about 100 times dimmer than the one associated with SGR 1935+2154 in April 2020.

Using SSE allows us to exploit the full potential of AMEGO-X and contribute to an exciting and successful mission.

% The AAS Journals would like to encourage authors to change software and
% third party data repository references from the current standard of a
% footnote to a first class citation in the bibliography.  As a bibliographic
% citation these important references will be more easily captured and credit
% will be given to the appropriate people.

% The first step to making this happen is to have the data or software in
% a long term repository that has made these items available via a persistent
% identifier like a Digital Object Identifier (DOI).  A list of repositories
% that satisfy this criteria plus each one's pros and cons are given at \break
% \url{https://github.com/AASJournals/Tutorials/tree/master/Repositories}.

% In the bibliography the format for data or code follows this format: \\

% \noindent author year, title, version, publisher, prefix:identifier\\

% \citet{2015ApJ...805...23C} provides a example of how the citation in the
% article references the external code at
% \doi{10.5281/zenodo.15991}.  Unfortunately, bibtex does
% not have specific bibtex entries for these types of references so the
% ``@misc'' type should be used.  The Repository tutorial explains how to
% code the ``@misc'' type correctly.  The most recent aasjournal.bst file,
% available with \aastex\ v6, will output bibtex ``@misc'' type properly.

%% IMPORTANT! The old "\acknowledgment" command has be depreciated. It was
%% not robust enough to handle our new dual anonymous review requirements and
%% thus been replaced with the acknowledgment environment. If you try to 
%% compile with \acknowledgment you will get an error print to the screen
%% and in the compiled pdf.

\begin{acknowledgments}
The material is based upon work supported by NASA under award number 80GSFC21M0002. Some of the results in this paper have been derived using the healpy and HEALPix package. This work has made use of the NASA Astrophysics Data System.
\end{acknowledgments}

\bibliography{references}{}
\bibliographystyle{aasjournal}

%% This command is needed to show the entire author+affiliation list when
%% the collaboration and author truncation commands are used.  It has to
%% go at the end of the manuscript.
%\allauthors

%% Include this line if you are using the \added, \replaced, \deleted
%% commands to see a summary list of all changes at the end of the article.
%\listofchanges

\end{document}